\documentclass[12pt]{iopart}
\usepackage{graphicx}
\expandafter\let\csname equation*\endcsname\relax
\expandafter\let\csname endequation*\endcsname\relax
\usepackage{amsmath}

\begin{document}
\bibliographystyle{unsrt}
\title[Effect of solar free oscillations on TianQin's range acceleration noise]{Effect of solar free oscillations on TianQin's range acceleration noise}

\author{Kun Liu, Chengjian Luo, and Xuefeng Zhang}

\address{MOE Key Laboratory of TianQin Mission, TianQin Research Center for Gravitational Physics \& School of Physics and Astronomy, Frontiers Science Center for TianQin, Gravitational Wave Research Center of CNSA, Sun Yat-sen University (Zhuhai Campus), Zhuhai 519082, China}
\ead{zhangxf38@sysu.edu.cn}
\vspace{10pt}


\begin{abstract}
TianQin is a proposed space-based gravitational-wave detector mission to be deployed and operated in high Earth orbits. As a sequel to [Zhang \textit{et al}. Phys. Rev. D 103, 062001 (2021)], we investigate a type of ``orbital noise'' in TianQin's range acceleration that is caused by gravitational perturbation associated with solar free oscillations. Frequencies of such oscillations are typically within TianQin's measurement band of 0.1 mHz--1 Hz, and the disturbance level needs careful assessment. By using high-precision orbit propagation and adding the Sun's time-variable oblateness $J_2$ to detailed gravity-field models, we examine the effect in the frequency domain and show that the solar free oscillation noise is expected to be two orders of magnitude lower than the noise requirement on single links and hence has little impact on the mission.
\end{abstract}

\section{Introduction} \label{sec:intro}

\begin{figure}
\centerline{\includegraphics[scale=0.15]{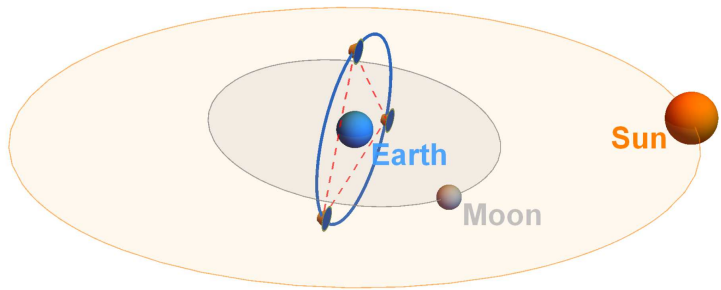}}
\caption{\label{fig:TQ} An illustration of the Sun, the Moon, and three TianQin satellites in an Earth-centered reference frame (figure reproduced from \cite{Ye2021}).}
\end{figure}

TianQin is designed to detect gravitational waves (GWs) in the millihertz frequency range (0.1 mHz--1 Hz), with a constellation consisting of three geocentric spacecrafts forming a nearly equilateral triangle set almost vertically to the ecliptic plane (see figure \ref{fig:TQ} and basic mission parameters in table \ref{tab:para}) \cite{Luo2016}. The detector uses laser interferometry to measure tiny distance variations between free-floating test masses inside each satellite \cite{LISA2000}, which can detect sources such as galactic ultra-compact binaries, stellar-mass black hole binaries, massive black hole binaries, extreme mass ratio inspirals, and stochastic GW background \cite{Luo2016, Mei2021}. 

The concept faces challenges from various environmental factors in space, including but not limited to gravity field, thermal environment, plasma, micrometeoroids, etc., among which the space gravity-field environment is of particular interest. The highly-sensitive inter-satellite ranging measurements can be influenced by the nearby gravity field. At large scales, the arm lengths of the constellation change with time, which requires orbit optimization to meet the configuration stability requirements \cite{Ye2019}. At small scales, it affects the geodesic motion of the test masses and also may interfere with GW signals in the detection frequency band.

A considerable number of interference sources from the Earth-Moon system have been evaluated in our previous works. Most of the orbital noises are in the frequency band below $10^{-4}$ Hz \cite{Zhang2021}, and how the effects vary with different choices of orbital orientations and radii has been discussed in \cite{Luo2022}. Nevertheless, we also need to discuss other potential sources which may enter the detection band. Here our work focuses on the solar free oscillation.

\begin{table}[ht]
\caption{\label{tab:para}TianQin basic mission parameters.}
\begin{indented}
\item[]\begin{tabular}{@{}cc}
\br
	Parameter&Value\\
\mr
	\hline
	Number of spacecraft&$N=3$\\
	Constellation&Equilateral triangle \\
	Type of orbits&Geocentric\\
	Arm length&$\sim 1.7\times 10^{5}$ km\\
	Position measurement noise&1 pm/Hz$^{1/2}$\\
	Residual acceleration noise&$1\times 10^{-15}$ m/s$^{2}$/Hz$^{1/2}$\\
\br
\end{tabular}
\end{indented}
\end{table}

Solar oscillations can generate long-term stable noise background overlapping with the frequency band of white dwarf binaries \cite{Cutler1996, Ni2000}. The work of \cite{Cutler1996} has estimated the magnitudes of the external gravitational perturbations that would be required for the solar oscillations to be observed by LISA \cite{Amaroseoane2017}. The estimates use the tidal acceleration formula with all-sky averaging and are based on observational upper limits and theoretical prediction of mode energies. To compare some published results, research on the proposed ASTROD mission has also evaluated some typical solar modes \cite{Ni2002}, and the strain amplitudes are estimated to be $10^{-25}$--$10^{-23}$, which are consistent with the results for LISA \cite{Cutler1996}. The work shows that ASTROD may have the ability to separate the solar oscillation signals from the GW background signals \cite{Ni2000}. The references \cite{Giampieri1999, Polnarev2009, Polnarev2009b} further investigated the prospects of observing combined signals of Newtonian gravity and GWs generated by solar quadrupole oscillations. Similarly for TianQin in geocentric orbits, we also need to evaluate the strength of these signals to see if these could affect GW detection.

The paper aims to simulate and evaluate the range-acceleration amplitude spectral density (ASD) generated by the solar free oscillation and compare them with the noise requirement of TianQin. Unlike previous works for LISA, we will use high precision numerical orbit simulation instead. The paper is structured as follows. In section \ref{sec:solosc}, we discuss the properties of the solar free oscillations, as well as the Newtonian and GW contributions to the range acceleration noise. In section \ref{sec:sim}, we use a high-precision orbit simulation program with various force models to obtain the expected noise spectrum of solar free oscillations, and make a comparison with TianQin's requirement. Finally, we present the conclusions in section \ref{sec:conclu}.

\section{Properties of the Solar Modes} \label{sec:solosc}
The oscillation in the solar atmosphere was first discovered in the 1960s \cite{Leighton1962}. Such photospheric periodic motions have a discrete frequency spectrum and have been observed in large numbers in the millihertz range \cite{Stix2002}. Large-scale projects such as SOHO (Solar and Heliospheric Observatory) \cite{Gabrial1991}, GONG (Global Oscillations Network Group) \cite{Harvey1988} and BiSON (The Birmingham Solar Oscillations Network) \cite{Elsworth1991} can obtain nearly continuous observations of the solar free oscillation. Millions of modes have been identified so far, and have helped us infer the acoustic and dynamical structure of the solar interior and place constraints on the models of the solar evolution \cite{Polnarev2009b}. However, the degree 2, low-order p, f, and g modes have not been observed so far with helioseismic velocity and intensity measurements, as its theoretically predicted amplitudes are far smaller than the threshold of current detectability \cite{Cutler1996, Polnarev2009, Gough1996, Kumar1996}.  These modes are closer to the denser part of the Sun's center, which suggests that space-based GW detectors like LISA may be more sensitive to these modes than conventional techniques \cite{Polnarev2009, Christensen1996}. 

One can describe the solar modes with the spherical harmonic function $Y_l^m(\theta ,\phi)$ and three indices $n$, $l$, $m$, which may determine the frequencies and lifetimes of solar oscillation modes \cite{Harvey1996}. Due to that the lifetimes of low-$l$, low-$n$ modes with low frequencies are expected to last at least several months or much longer \cite{Harvey1996}, we regard the low-$l$ and low-$n$ solar oscillation noises as approximately stationary in one observation window (3 months) of TianQin \cite{Luo2016}. Among the millions of different solar oscillation modes, we mostly care about the quadrupole modes ($l=2$) that can generate largest perturbations to the gravitational potential \cite{Giampieri1999, Polnarev2009}. In addition, lower-order modes excite significantly larger external gravitational perturbations than higher-order modes given fixed mode energy \cite{Cutler1996}. Therefore, we can determine that the low-$n$ quadrupole modes are more relevant to TianQin.

These modes with low frequencies are expected to have energies about ${10^{28}}$ ergs \cite{Cutler1996, Libbrecht1991, Ni2002}. The associated quadrupole moments $\alpha$ are given in \cite{Cutler1996} (see table \ref{tab:modes}), with mode energies normalized to ${10^{28}}$ ergs. Other sources on the values of $\alpha$ can be seen in table \ref{tab:Ni} and \ref{tab:uppperlimit}. The gravitational perturbation is expressed as 
\begin{eqnarray}\label{equ:culter}
\begin{aligned}
	\delta \Phi  = \alpha \frac{G{M_ \odot }}{R_ \odot}{\left(\frac{{{R_ \odot }}}{r}\right)^{l + 1}}{Y_l^m(\theta ,\phi)}{e^{i\omega t}}, 
	\end{aligned}
\end{eqnarray}

where $M_\odot$ is the Sun's mass, $R_\odot$ is the Sun's radius, $\omega$ is the oscillation angular frequency, and $r$ is the distance from the Sun to the satellite \cite{Cutler1996}. Due to the factor $(1/r)^{l+1}$ and that the TianQin satellites are located at a distance of about 1 AU from the Sun, the gravitational perturbation decreases rapidly with increasing degree $l$. Based on Eq. (\ref{equ:culter}), the corresponding strain amplitudes can be calculated from the tidal acceleration equation \cite{Cutler1996}.

\begin{table}[t]
\caption{\label{tab:modes}Parameters of the solar modes used in the simulation \cite{Cutler1996}. The corresponding strain amplitudes $h$ are also included in the table. For LISA, the detectable strain amplitude is estimated to be ${10^{-21}}$--${10^{-20}}$ at a signal-to-noise ratio of 5 with one year's observation \cite{Polnarev2009}.}
\begin{indented}
\item[]\begin{tabular}{@{}cccc}
\br
	Mode&Frequency ($\mu$Hz)&$\alpha$&$h$\\
\mr
	\hline
            g7&$101.2$&$1.42\times{10^{-12}}$&$1.10\times{10^{-23}}$\\
            g6&$134.0$&$2.80\times{10^{-12}}$&$1.24\times{10^{-23}}$\\
            g5&$168.8$&$4.72\times{10^{-12}}$&$1.32\times{10^{-23}}$\\
            g4&$192.2$&$5.96\times{10^{-12}}$&$1.29\times{10^{-23}}$\\
            g3&$220.4$&$7.17\times{10^{-12}}$&$1.18\times{10^{-23}}$\\
            g2&$254.0$&$6.07\times{10^{-12}}$&$7.50\times{10^{-24}}$\\
            g1&$293.6$&$2.81\times{10^{-12}}$&$2.60\times{10^{-24}}$\\
            f&$350.9$&$2.31\times{10^{-12}}$&$1.50\times{10^{-24}}$\\
            p1&$381.6$&$5.31\times{10^{-12}}$&$2.91\times{10^{-24}}$\\
            p2&$514.4$&$2.02\times{10^{-12}}$&$6.07\times{10^{-25}}$\\
            p3&$663.6$&$8.60\times{10^{-13}}$&$1.56\times{10^{-25}}$\\
\br
\end{tabular}
\end{indented}
\end{table}

\begin{table}[t]
\caption{\label{tab:Ni}The quadrupole moments estimated from limiting the velocity amplitudes $V$, which are converted from \cite{Ni2000,Ni2002}. When the effective bulk amplitudes $\bar\xi _{nlm}=1$ mm, the quadrupole moments are about $3.61\times{10^{-12}}$.}
\begin{indented}
\item[]\begin{tabular}{@{}cccc}
\br
		Mode&Frequency ($\mu$ Hz)&$\alpha$ ($V=1$ mm/s)&$h$\\
\mr
	\hline
        g8&$121.1$&$3.51\times{10^{-11}}$&$9.71\times{10^{-23}}$\\
        g7&$134.0$&$3.41\times{10^{-11}}$&$9.43\times{10^{-23}}$\\
        g6&$149.8$&$3.23\times{10^{-11}}$&$8.93\times{10^{-23}}$\\
        g5&$168.8$&$3.09\times{10^{-11}}$&$8.55\times{10^{-23}}$\\
        g4&$192.2$&$2.62\times{10^{-11}}$&$7.25\times{10^{-23}}$\\
        g3&$220.4$&$1.97\times{10^{-11}}$&$5.46\times{10^{-23}}$\\
        g2&$254.0$&$1.17\times{10^{-11}}$&$3.25\times{10^{-23}}$\\
        g1&$293.6$&$3.87\times{10^{-12}}$&$1.07\times{10^{-23}}$\\
        f&$350.9$&$2.30\times{10^{-12}}$&$6.37\times{10^{-24}}$\\
        p1&$381.6$&$3.14\times{10^{-12}}$&$8.70\times{10^{-24}}$\\
        p2&$514.4$&$4.41\times{10^{-13}}$&$1.22\times{10^{-24}}$\\
        p3&$663.6$&$1.01\times{10^{-13}}$&$2.80\times{10^{-25}}$\\
\br
\end{tabular}
\end{indented}
\end{table}

\begin{table}[t]
\caption{\label{tab:uppperlimit}Quadrupole moments with observational upper limits of 1 cm/s on solar surface velocity amplitudes, which are converted from \cite{Polnarev2009}.}
\begin{indented}
\item[]\begin{tabular}{@{}cccccc}
\br
		Mode&Frequency ($\mu $Hz)&$\alpha$&Mode&Frequency ($\mu $Hz)&$\alpha$\\
\mr
	\hline
            g22&50.6&$1.43\times{10^{-11}}$&g7&136&$1.46\times{10^{-10}}$\\
            g21&52.9&$1.60\times{10^{-11}}$&g6&152&$1.74\times{10^{-10}}$\\
            g20&55.4&$1.77\times{10^{-11}}$&g5&171&$2.00\times{10^{-10}}$\\
            g19&58.1&$2.00\times{10^{-11}}$&g4&194&$2.12\times{10^{-10}}$\\
            g18&61.1&$2.28\times{10^{-11}}$&g3&222&$2.00\times{10^{-10}}$\\
            g17&64.4&$2.62\times{10^{-11}}$&g2&257&$1.46\times{10^{-10}}$\\
            g16&68.0&$3.00\times{10^{-11}}$&g1&297&$7.19\times{10^{-11}}$\\
            g15&72.1&$3.46\times{10^{-11}}$&f&356&$5.28\times{10^{-11}}$\\
            g14&76.8&$4.06\times{10^{-11}}$&p1&384&$7.19\times{10^{-11}}$\\
            g13&82.0&$4.81\times{10^{-11}}$&p2&515&$1.48\times{10^{-11}}$\\
            g12&87.9&$5.70\times{10^{-11}}$&p3&664&$4.52\times{10^{-12}}$\\
            g11&94.7&$6.81\times{10^{-11}}$&p4&812&$1.58\times{10^{-12}}$\\
            g10&103&$8.29\times{10^{-11}}$&p5&960&$6.24\times{10^{-13}}$\\
            g9&112&$1.00\times{10^{-10}}$&p6&1110&$2.71\times{10^{-13}}$\\
            g8&123&$1.22\times{10^{-10}}$& & &\\
            
\br
\end{tabular}
\end{indented}
\end{table}

In order to conduct orbit simulation by modifying the gravity field coefficients in the program we developed, we rewrite gravity potential change $\delta\Phi$ at radius $r$, co-latitude $\theta$ and longitude $\phi$ in the form of equation (\ref{equ:Montenbruck}) in \cite{Montenbrunk2000}, which is often seen in physical geodesy:
\begin{eqnarray}\label{equ:Montenbruck}
    \delta\Phi =\frac{GM_\odot}{R_ \odot}\sum\limits_{l=0}^\infty {\left(\frac{{{R_ \odot }}}{r}\right)^{l+1}}\sum\limits_{m=0}^l(\delta C_{lm}\cos(l\phi) + \delta S_{lm}\sin(l\phi))P_{lm}(\cos(\theta)), 
\end{eqnarray}
where $P_{lm}$ is the associated Legendre function of spherical harmonic degree $l$ and order $m$. Gravity potential coefficients $\delta C_{lm}$ and $\delta S_{lm}$ describe the dependence on the time-varying Sun's internal mass distribution and also represent changes of the solar gravity field. Since $ C_{20}=-J_{2}$, the coefficient $\delta C_{20}$ with corresponding oscillation frequencies can be obtained from the quadrupole moments of the solar modes in table \ref{tab:modes}.

The solar oscillations can also generate GWs. The effect dominates in the frequency band above ${\nu _r} \equiv c/2\pi r \approx$ 0.3 mHz, and is larger than the Newtonian counterpart by a factor of the order $(2\pi r \nu/c)^4$ \cite{Polnarev2009, Polnarev2009b}. The contribution of GWs will also be presented in the next section. 
\normalsize


\section{Simulation setup and results} \label{sec:sim}
We adopt the range acceleration $\ddot \rho$ as the observable, which is the second time derivative of the range $\rho$ between two satellites, and compare the ASD with the residual acceleration noise requirements ($1.4\times 10^{-15}$ m/s$^{2}$/Hz$^{1/2}$). To meet the precision requirement, our team have developed a package TQPOP (TianQin Quadruple Precision Orbit Propagator) and extended the precision to 34 significant digits with quadrupole precision arithmetic \cite{Zhang2021}. The detailed force models can be found in table 2 in \cite{Luo2022} and the subsequent simulation is to further consider the perturbations of the Sun's $J_{2}$ given in section \ref{sec:solosc} and analyze their impact on the TianQin's detection.

The initial orbital parameters from table 1 in \cite{Luo2022} are adopted as input and the integration lasts for one observation window of 3 months \cite{Luo2016}. In order to highlight the contribution of the solar free oscillation to the total effect, we only retain some necessary force models in the initial simulations, such as the point mass of the Earth and the Sun, and do not consider models like EOP (Earth Orientation Parameters), so that the numerical errors of high-frequency parts can be significantly reduced.

\begin{figure}[htb]
	\centering
	\includegraphics[scale=0.7]{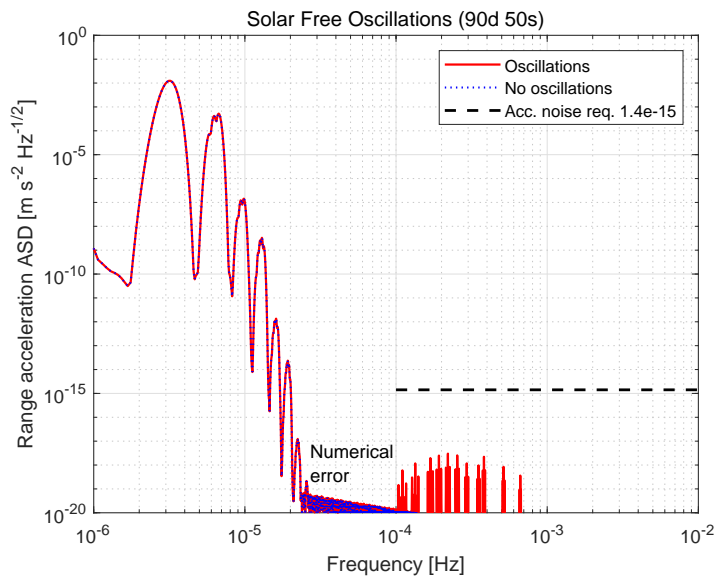}
	\caption{\label{fig:osc}The range-acceleration ASD of two TianQin satellites after 90 days with and without the solar free oscillations, using the parameters from table \ref{tab:modes}.}
\end{figure}

\begin{figure}
	\centering
	\includegraphics[scale=0.7]{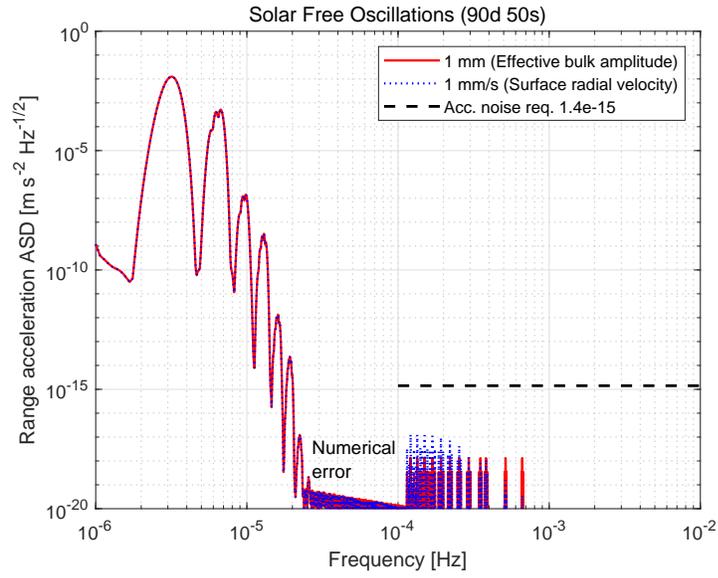}
	\caption{\label{fig:Ni}The contributions of solar free oscillations to the range-acceleration ASD from two different estimation methods for mode amplitudes (see table \ref{tab:Ni}).}
\end{figure}

In figure \ref{fig:osc}, we simulate the range acceleration of two satellites with and without the solar free oscillations (only modes above $10^{-4}$ Hz are considered). The splitting of the peaks is caused by coupling with the orbital frequency ($3\times{10^{-6}}$ Hz), which corresponds to an orbital period of about 3.64 days. In this test case, the peaks at $<10^{-4}$ Hz are due to the point masses of the Earth and the Sun. As a comparison, we also present the simulation using the parameters from table \ref{tab:Ni} \cite{Ni2000} in figure \ref{fig:Ni}, which also meets the range acceleration noise requirement and is close to the magnitude in figure \ref{fig:osc}.

\begin{figure}[htb]
\centering
\includegraphics[scale=0.7]{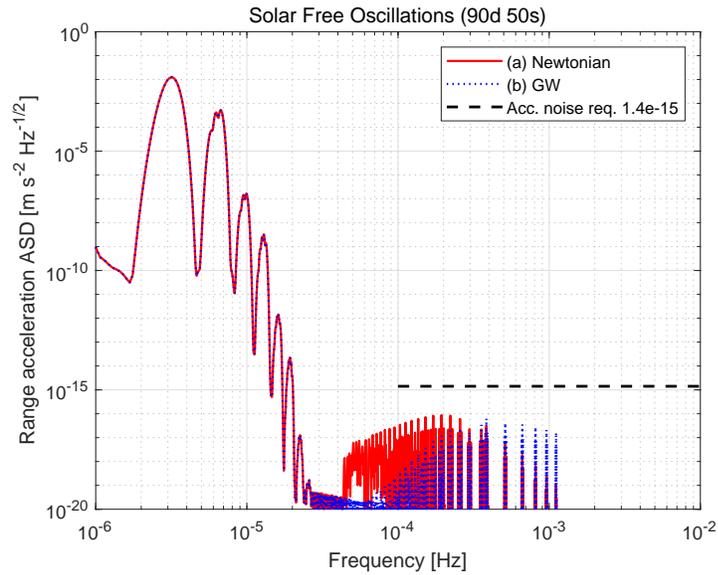}
\caption{\label{fig:upperlimits} Quadrupole modes with observational upper limits of 1 cm/s on solar surface velocity amplitudes (see table \ref{tab:uppperlimit}). Curve (a) represents Newtonian part and (b) represents GW part.}
\end{figure}

Since the above parameters of modes are theoretical predictions, we also need to give the maximum estimation based on the actual observational upper limit. If the upper limit of the surface velocity amplitude 1 cm/s is considered \cite{Polnarev2009, Appourchaux2010}, the worst case can be simulated as shown by the red curve in figure \ref{fig:upperlimits}, which is 1--2 orders higher than the previous results, but still lower than the range acceleration noise requirement. The contribution of emitted GWs to the noise is shown by the blue dotted line in figure \ref{fig:upperlimits}. In the frequency band above 0.3 mHz, the contribution of GW part is higher than the Newtonian part, but both are expected to decrease at the higher frequency band.
	
\begin{figure}[htb]
	\centering
	\includegraphics[scale=0.7]{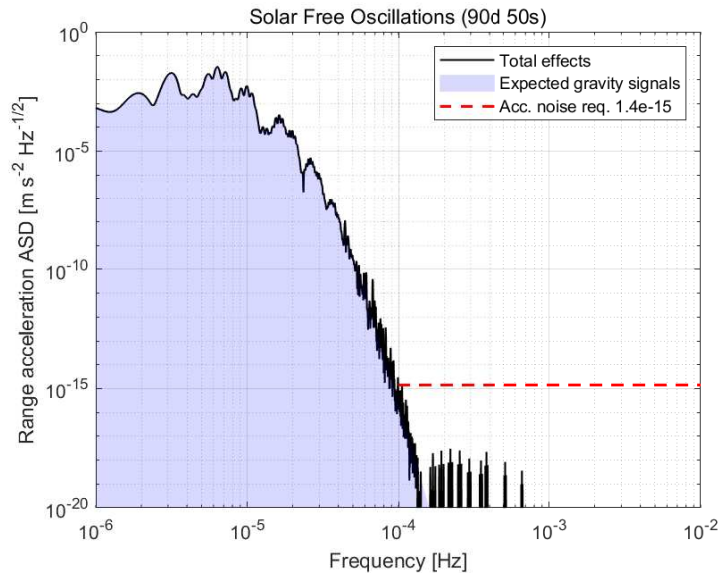}
	\caption{\label{fig:all} The total effects combining the solar free oscillations and the Earth-Moon system. }
\end{figure}

After adding other force models \cite{Luo2022} of the Earth-Moon system to the simulation of figure \ref{fig:osc}, we can see the total effects including the solar free oscillations in figure \ref{fig:all}. The result still meets the noise requirement of the range acceleration and compliments our previous result shown in figure 2 of \cite{Zhang2021}.

\section{Conclusion} \label{sec:conclu}

In this work, we have evaluated the disturbance caused by solar free oscillations and compared it with TianQin's noise requirement. Unlike previous works that have been done for other missions \cite{Cutler1996, Ni2000, Polnarev2009}, we have directly calculated their effects on inter-satellite range acceleration via high-precision orbit simulation. Given the current observation and theoretical estimates on the mode amplitudes, the solar oscillation noise is expected to be well below TianQin's total noise requirement by at least 1 order of magnitude. Therefore, we conclude that the solar free oscillation will not affect TianQin's GW detection.

Nevertheless, the work on evaluating gravity field disturbance does not end here for TianQin. Free oscillation frequencies of the Earth can also enter the target band of TianQin. These are currently being investigated and will be reported in the future.

\section*{Acknowledgment}
The authors thank Bobing Ye, Yuzhou Fang, Junwei Zhao, Cong Yu, Jianwen Ou, Jianwei Mei, Yiming Hu and Jiandong Zhang for their helpful discussions. X. Z. is supported by the National Key R\&D Program of China (Grant No. 2020YFC2201202). 

\section*{References}
\bibliography{apscite.bib}

\end{document}